\documentclass[%
 reprint,
 superscriptaddress,
 amsmath,amssymb,
 aps,
 pra,
floatfix,
]{revtex4-2}
\usepackage{graphicx}
\usepackage{dcolumn}
\usepackage{bm}
\usepackage[english]{babel}
\usepackage[utf8]{inputenc}
\usepackage{amsthm}
\usepackage{mathtools}
\usepackage{physics}
\usepackage{xcolor}
\usepackage{graphicx}
\usepackage[left=23mm,right=13mm,top=35mm,columnsep=15pt]{geometry} 
\usepackage{adjustbox}
\usepackage{placeins}
\usepackage[T1]{fontenc}
\usepackage{lipsum}
\usepackage{csquotes}
\usepackage{float}
\usepackage{hyperref}
\usepackage{enumitem}
\usepackage{appendix}

\usepackage{physics}
\usepackage{braket}
\usepackage{esvect}
\usepackage{amsmath}
\usepackage[color=yellow]{todonotes}
\begin{document}

\preprint{APS/123-QED}

\title{Relativistic VQE calculations of molecular electric dipole moments on quantum hardware}

\author{Palak Chawla}
\altaffiliation{Contributed equally to the work}
\affiliation{Centre for Quantum Engineering, Research and Education, TCG Crest, Kolkata 700091, India}
\author{Shweta}
\altaffiliation{Contributed equally to the work}
\affiliation{Department of Physics, Indian Institute of Technology Delhi, New Delhi 110016, India}
\author{K. R. Swain}
\altaffiliation{Current address: Department  of  Physics  and  Materials  Science,  University  of  Luxembourg,  L-1511  Luxembourg,  Luxembourg}
\affiliation{Centre for Quantum Engineering, Research and Education, TCG Crest, Kolkata 700091, India}
\author{Tushti Patel}
\affiliation{Centre for Quantum Engineering, Research and Education, TCG Crest, Kolkata 700091, India}
\affiliation{Department of Physics, IIT Tirupati, Chindepalle, Andhra Pradesh 517619, India}
\author{Renu Bala}
\altaffiliation{Current address: Institute of Physics, Faculty of Physics, Astronomy and Informatics, Nicolaus Copernicus University, Grudziadzka 5, 87-100 Toru\'n, Poland}
\affiliation{Centre for Quantum Engineering, Research and Education, TCG Crest, Kolkata 700091, India}
\author{Disha Shetty}
\affiliation{Centre for Quantum Engineering, Research and Education, TCG Crest, Kolkata 700091, India}
\author{Kenji Sugisaki}
\affiliation{Centre for Quantum Engineering, Research and Education, TCG Crest, Kolkata 700091, India}
\affiliation{Graduate School of Science and Technology, Keio University, 7-1 Shinkawasaki, Saiwai-ku, Kawasaki, Kanagawa 212-0032, Japan}
\affiliation{Quantum Computing Center, Keio University, 3-14-1 Hiyoshi, Kohoku-ku, Yokohama, Kanagawa 223-8522, Japan}
\affiliation{Keio University Sustainable Quantum Artificial Intelligence Center (KSQAIC), Keio University, 2-15-45 Mita, Minato-ku, Tokyo, Japan}
\author{Sudhindu Bikash Mandal}
\affiliation{Centre for Quantum Engineering, Research and Education, TCG Crest, Kolkata 700091, India}
\author{Jordi Riu}
\affiliation{Qilimanjaro Quantum Tech, Carrer de Veneçuela, 74, Sant Martí, 08019, Barcelona, Spain}
\affiliation{Universitat Politècnica de Catalunya, Carrer de Jordi Girona, 3, 08034 Barcelona, Spain}
\author{Jan Nogué}%
\affiliation{Qilimanjaro Quantum Tech, Carrer de Veneçuela, 74, Sant Martí, 08019, Barcelona, Spain}
\author{V. S. Prasannaa}
\email{srinivasaprasannaa@gmail.com}
\affiliation{Centre for Quantum Engineering, Research and Education, TCG Crest, Kolkata 700091, India}
\affiliation{Academy of Scientific and Innovative Research (AcSIR), Ghaziabad- 201002, India}
\author{B. P. Das}
\affiliation{Centre for Quantum Engineering, Research and Education, TCG Crest, Kolkata 700091, India}
\affiliation{Academy of Scientific and Innovative Research (AcSIR), Ghaziabad- 201002, India}
\affiliation{Department of Physics, Tokyo Institute of Technology, 2-12-1 Ookayama, Meguro-ku, Tokyo 152-8550, Japan}

\date{\today}

\begin{abstract} 
The quantum--classical hybrid variational quantum eigensolver (VQE) algorithm is among the most actively studied topics in atomic and molecular calculations on quantum computers, yet few studies address properties other than energies or account for relativistic effects. This work presents high-precision 18-qubit relativistic VQE simulations for calculating the permanent electric dipole moments (PDMs) of alkaline earth metal monohydride molecules (BeH to RaH) on traditional computers, and 6- and 12-qubit PDM computations for SrH on IonQ quantum devices. To achieve high precision on current noisy intermediate scale era quantum hardware, we apply various resource reduction methods, including reinforcement learning-based and causal flow preserving ZX--Calculus routines, along with error mitigation and post-selection techniques. Our approach reduces the two-qubit gate count in our 12-qubit circuit by $99.71\%$, with only a $2.35\%$ trade-off in precision for PDM when evaluated classically within a suitably chosen active space. On the current generation IonQ Forte-I hardware, the error in PDM is $-1.17\%$ relative to classical calculations and only $1.21\%$ compared to the unoptimized circuit. 
\end{abstract} 

\maketitle


\section{Introduction} 

The quantum--classical hybrid Variational Quantum Eigensolver (VQE) algorithm, which is built on the Rayleigh--Ritz variational principle, is the leading approach to calculating energies in the Noisy Intermediate Scale Quantum (NISQ) era \cite{Peruzzo2014VQE,guo2024experimental,Pan2023vqe12q, google2023purification, yamamoto2022quantum, eddins2022doubling, gao2021applications, kawashima2021optimizing, google2020hartree, nam2020ground, kandala2019error, mccaskey2019quantum, hempel2018quantum, colless2018computation, kandala2017hardware, o2016scalable, Jules2022VQE_review}. VQE is an iterative algorithm that involves minimizing an energy functional through energy evaluations on a quantum device and parameter updates on a traditional computer, and such a procedure yields an upper bound to the true ground state energy~\cite{gould1966vqe}. In particular, given a Hamiltonian, $\hat{H}$, and a suitably parametrized state, $\ket{\Psi(\theta)}$, where $\{\theta\}$ $\in$ $\{\theta_1,\theta_2, \ldots\}$, minimizing an energy functional $E(\theta) = \frac{\expval{\hat{H}}{\Psi(\theta)}}{\braket{\Psi(\theta)|\Psi(\theta)}} = \langle \Phi |U^\dag(\theta) \hat{H} U(\theta)|\Phi \rangle$ with respect to $\{\theta\}$ yields an optimized state, $\ket{\Psi(\theta^*)}$, such that $\langle \Phi |U^\dag(\theta^*) \hat{H} U(\theta^*)|\Phi \rangle \geq E_0$. Here, the state $\ket{\Psi(\theta)}$ is expressed as a unitary, $U(\theta)$ acting on a reference state (which for our purposes is the Hartree--Fock/Dirac--Fock (DF) state), $\ket{\Phi}$. The VQE approach, by construction, leads to relatively low-depth quantum circuits~\cite{Peruzzo2014VQE} when compared with quantum phase estimation, for example, thus making it the workhorse for atomic and molecular calculations in the NISQ era. 

\begin{figure}[t]
\centering
\setlength{\tabcolsep}{1mm}
\begin{tabular}{c}
\hspace{-1.1cm}
\includegraphics[height=90mm,width=90mm]{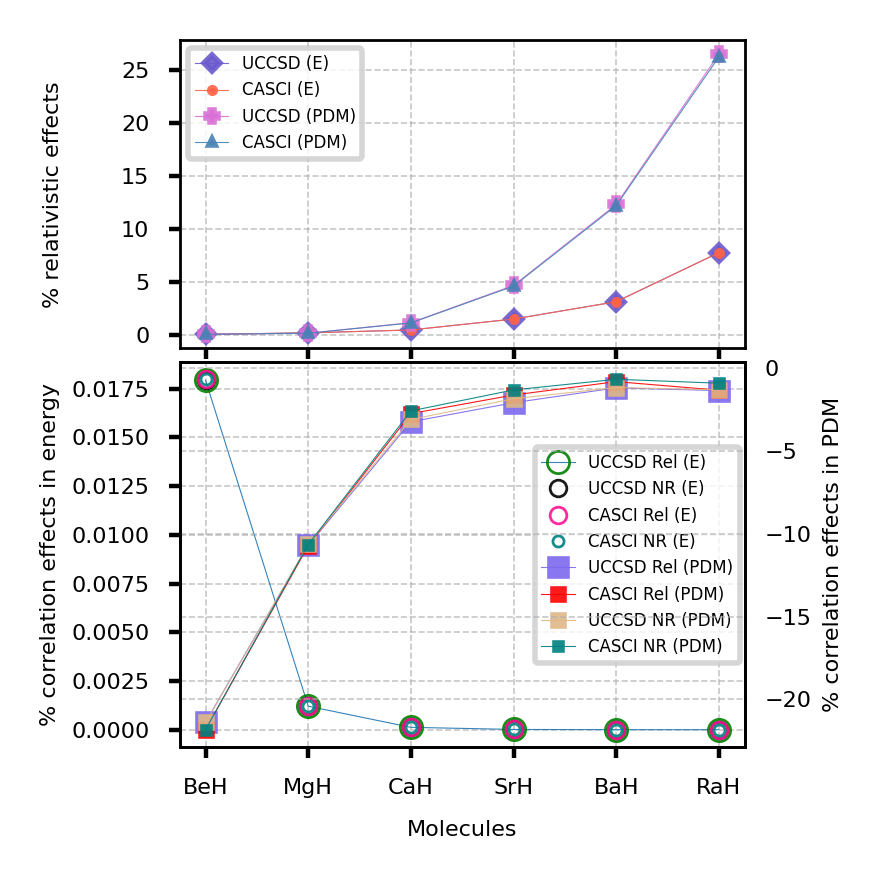} 
\vspace{-0.8cm}
\end{tabular}
\caption{The top panel shows the percentage relativistic effects = $\frac{A_{Rel}-A_{NR}}{A_{Rel}} \times 100$ for ground state energy (E) and PDM from our 18-qubit VQE simulations. Our results are benchmarked against CASCI calculations. The bottom panel shows the percentage correlation effects = $\frac{A_{X}-A_{MF}}{A_{X}} \times 100$, where $X$ can be correlation energy (circular markers) or the PDM (square markers) relative to respective quantity at VQE and CASCI levels, and MF refers to the result obtained at Mean Field level of theory. } 
\label{fig:1}
\end{figure} 

Despite its successes, the algorithm has been limited in its scope of applications: it has mainly been employed to calculate ground state energies, ionization energies, and excitation energies \cite{Ostaszewski2021optimisation_qc, Tang2021AdaptVqe,Villela2022sim_ionizationenergy, sugisaki2021quantum, Sumeet2022precision_vqe,Bharti2022NISQalgo,Mario2021QCalgos_chemistry}, whereas VQE's application to other properties, such as the molecular permanent electric dipole moments (PDMs), is limited \cite{Stober2022thermo_propert, Parrish2019ele_trans_vqe, Villela2022sim_ionizationenergy}. There is also little in VQE literature on extending it to the relativistic regime \cite{Zaytsev2022moscovium_gs, swain2022molecular}. In fact, there are only a handful of studies that incorporate relativistic effects in the context of even other quantum algorithms \cite{Veis2012rel_qchemistry, Stetina2022QED_qc, Sugisaki2023BPD_algo, vikrant2023rel_QA}. The inclusion of relativity in atomic and molecular physics calculations has lead to several notable predictions, including explaining the colour of gold~\cite{pyykko1979relativity}, relativistic contraction of bond lengths~\cite{pyykko1979relativity}, and the rationale behind starting of cars with lead-acid batteries \cite{ahuja2011relativity}. A vast landscape of atomic and molecular properties where relativity is important, ranging from practically useful atomic clocks to probing particle physics \cite{chupp2019electric,yamanaka2017probing, ohayon2022nuclear, ohayon2024reconciling, sahoo2021new, tang2023simultaneous, safronova2018search, prasannaa2015mercury, sunaga2019ultracold, pavsteka2017relativistic, borschevsky2013relativistic, haase2020hyperfine}, remains to be explored in the quantum computing framework. Thus, it is timely and relevant to: a. to calculate properties other than energies as here too, there is little available work in literature although energies are but very few of the vast landscape of atomic and molecular properties, and b. carry out relativistic VQE calculations given that the algorithm is widely used and yet has a notable gap in literature with almost no focus on a plethora of applications involving relativistic effects. 

It is important to note that there is nothing in VQE that inherently limits it from dealing with the PDMs and other wavefunction-based quantities. Furthermore, the same can be said of its ability to handle relativistic effects. Both of these aspects, in fact, depend on the ability to carry out classical pre-processing, that is, obtain the PDM integrals and the one- and two- body integrals, all in the relativistic framework. The dearth of literature on this topic is likely due to three reasons. The first is the specialized nature of incorporating relativistic effects into quantum many-body theoretic computations. It is also important to add that the number of programs that can generate the required Hamiltonian integrals and property integrals with the Dirac-Coulomb Hamiltonian are far fewer than those for non-relativistic calculations. This is likely the second reason why fully relativistic quantum chemical computations have not yet been adopted by the quantum computing community, as it involves a `learning curve'. Lastly, since quantum computing is itself still a relatively new field and the computational resources required for carrying out calculations are still expensive, it is not surprising that many important problems such as property calculations in the relativistic framework are yet to be explored. 

The molecular property chosen for this study  is the PDM, due to its usefulness in several applications, including the search for novel quantum phases like the supersolid phase \cite{Tapan2009BoseHubbard}, study of dipole--dipole molecular interactions with implications in quantum computing \cite{Norrgard2016RFtrap}, and fundamental physics searches \cite{Amar2010edm}. In particular, the PDMs of the alkaline earth metal monohydrides are of significant interest due to their potential for laser cooling \cite{Gao2014lc, Pang2023lc}. Notably, CaH and BaH have already been laser-cooled \cite{McNally_2020, Carson2022CaH}. 

We now move to our third motivation behind the work. Despite the limitations of quantum hardware in the NISQ era, it is crucial to push the boundaries of the current best available technology. Doing so not only underscores the progress achieved, but also offers valuable insights for advancing and guiding the development of quantum algorithms as well as quantum hardware in the near-term. To that end, we compute the PDM of the moderately heavy SrH molecule in a 12-qubit relativistic VQE calculation on among the best available commercial quantum computers, the current generation IonQ Forte-I device. Typical VQE calculations on quantum hardware using the physics/chemistry-inspired unitary coupled cluster (UCC) ansatz employ resource reduction strategies to minimize qubits, gates, and VQE iterations, along with error mitigation techniques, to predict energies~\cite{guo2024experimental, shang2024rapidly, ShuoSun2024gs, dahale2023quantum, Zhao2022ooupccd, Sarma2023neutron_qc, zhao2023aria, Ollitrault2020eom}. The largest such computation carried out to date involved 12-qubits~\cite{guo2024experimental}, but the framework was non-relativistic and the property of interest was the ground state energy. In this work, we conduct 12-qubit relativistic VQE computations of the PDM, but due to the deep circuits incurred for such tasks, a significant part of our effort focuses on resource reduction. This allows us to achieve sufficiently shallow circuit depths for attaining reasonable precision, while carefully preserving the underlying physics to the best capabilities of NISQ computers. We add that a relativistic treatment necessitates an \textit{additional} need for aggressive resource reduction strategies, since it generates significantly more integrals compared to the non-relativistic case, thus leading to evaluating more quantum circuits. In the NISQ era, even a modest increase in the additional number of required circuits can place a severe strain on the quality of results. 

\begin{figure*}[t]
\setlength{\tabcolsep}{0.6mm}
        \begin{tabular}{cc}
        \hspace{-0.6cm}
            \centerline{\includegraphics[scale=0.8]{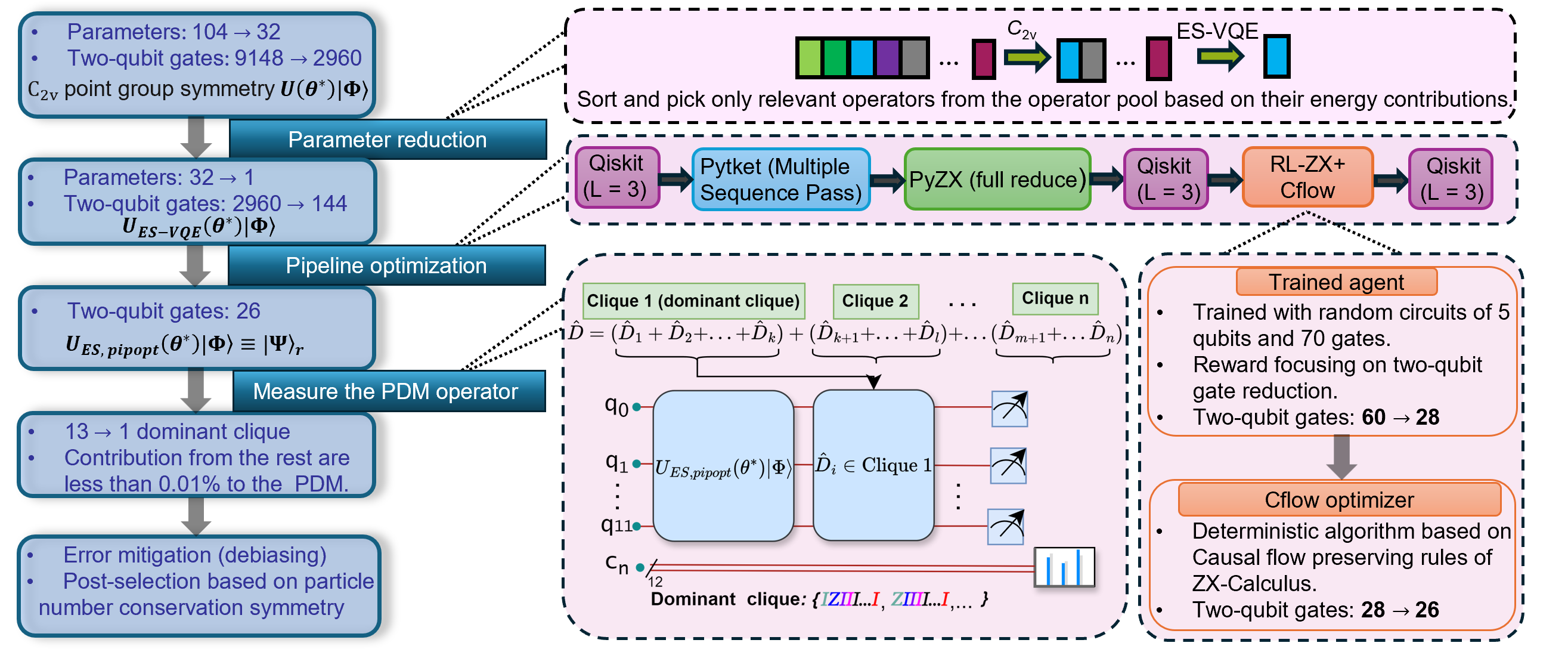}}
            \\
            \hspace{0.3cm}(a)\\
    
        \end{tabular}
        \begin{tabular}{ccc}
        \hspace{-0.5cm}
            \includegraphics[height=40mm,width=62mm]{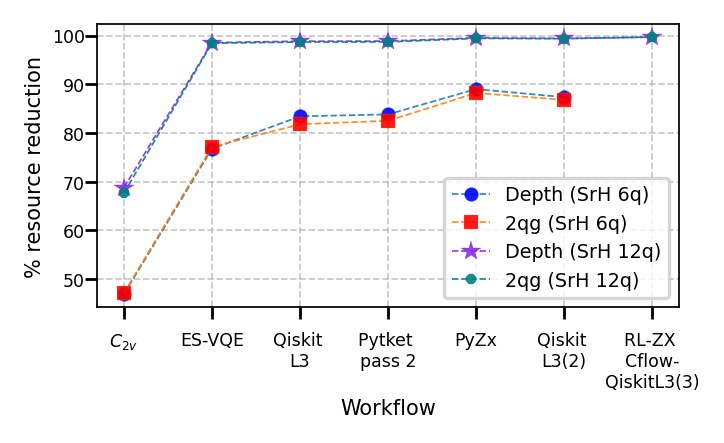}&
            \hspace{-0.4cm}
            \includegraphics[height=40mm,width=62mm]{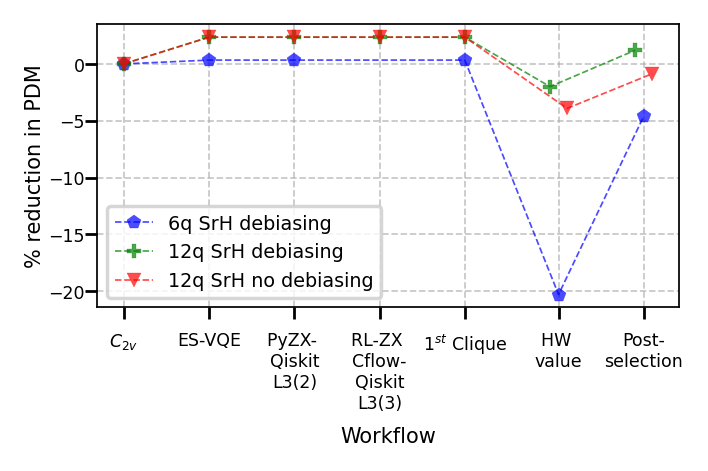}&
            \hspace{-0.5cm} \raisebox{2.5pt}
            {\includegraphics[height=40mm, width=62mm]{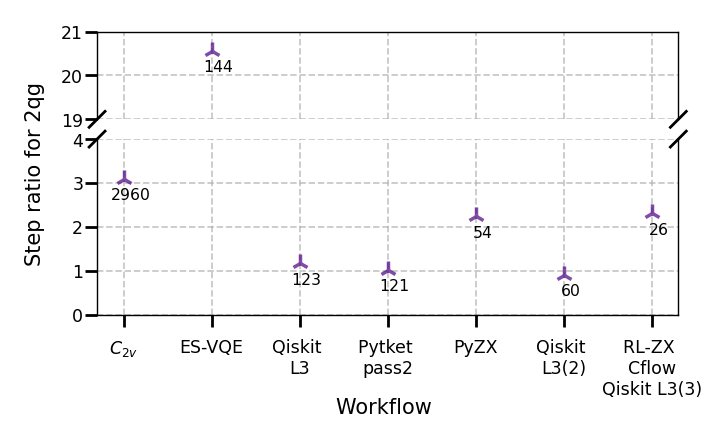}}\\
            (b)&\hspace{1cm}(c)&\hspace{1cm}(d)\\
        \end{tabular}
\caption{(a) Our workflow for quantum hardware execution of SrH 12-qubit PDM calculation on the IonQ Forte device, which leads to reducing quantum resources while retaining precision. (b) Percentage reduction in resources (two-qubit gates, denoted as $2qg$ in the sub-figure, and circuit depth) with each step of our workflow: $U_{ES\mathchar`-VQE}$ is the UCCSD circuit post-ES-VQE and $U_{ES,Pipopt}$ is the state after pipeline-based optimization (denoted as pipopt). (c) The loss of precision in predicting PDM after each step in our workflow, with `1$^{st}$ Clique' indicating the selection of the dominant clique for the PDM operator (See Table \ref{SMtable3} of the Supplemental Material). Sub-figure (d) illustrates the step ratio, which is the ratio of the number of $2qg$ before and after the current step in our workflow. It is important to stress that the compound strategy of our RL--ZX based agent followed by the causal flow deterministic algorithm (both based on ZX--Calculus, denoted as RL--ZX $+$ Cflow) reduces the already small gate count by one half. } 
\label{fig:2}
\end{figure*} 

\section{Theory and Methodology}

We now discuss some details of the VQE algorithm to calculate the PDM. The quality of a VQE calculation is predominantly determined by choice of $U(\theta)$. We choose the UCC ansatz for $U(\theta)$, which being both variational and unitary, is well-suited for quantum computers while retaining the predictive accuracy of traditional CC methods, which are deemed to be the gold standard of electronic structure calculations~\cite{Abe2014rel_eff_EE}. The UCC ansatz is given by $\ket{\Psi(\theta)} = e^{\hat{T}-\hat{T}^\dagger} \ket{\Phi}= e^{\hat{\tau}} \ket{\Phi}$. The ansatz includes single ($\hat{T_1}$) and double ($\hat{T_2}$) excitations (UCCSD approximation): $\hat{T_1}=\sum_{ia} \theta^a_i \hat{a}_a^\dagger \hat{a}_i$ , $\hat{T_2} = \sum_{ijab} \theta^{ab}_{ij} \hat{a}_a^\dagger \hat{a}_b^\dagger \hat{a}_j \hat{a}_i$. $i,j,\cdots$ indicate indices of occupied spinorbitals, while $a,b,\cdots$ are those of unoccupied/virtual ones. We note that the Hamiltonian is given by $\hat{H} = \sum_{pq} h_{pq} \hat{a}_p^\dagger \hat{a}_q + \frac{1}{2} \sum_{pqrs} h_{pqrs} \hat{a}_p^\dagger\hat{a}_q^\dagger \hat{a_s}\hat{a_r}$ in the second quantized notation, where $h_{pq}$ and $h_{pqrs}$ are the one- and two-electron integrals, and $p, q,\cdots$ can denote indices of occupied or unoccupied spinorbitals. Upon using the Jordan--Wigner transformation~\cite{Seeley2012The} for both the state and the Hamiltonian, the energy functional can be expressed as $E(\theta) = \sum_{m}^{M} \alpha_m \expval{U'^{\dagger}(\theta) \hat{P}_m U'(\theta) }{\Phi'}$. $U'(\theta)$ and $\ket{{\Phi'}}$ refer to the Jordan--Wigner-transformed $U(\theta)$ and $\ket{{\Phi}}$, respectively. $\alpha_m$ are the coefficients depending on $h_{pq}$ and $h_{pqrs}$ and $P_m$ are Pauli strings represented by tensor products of Pauli operators $\{I,X,Y,Z\}$. Each term in the equation corresponds to the expectation value of a Pauli string $\hat{P}_m$, which is evaluated through statistical sampling on a quantum computer, while the summation and energy minimization is done on a traditional computer. Once the optimized parameters are determined, we evaluate $\sum_{n}^{N} d_n \expval{U'^{\dagger}(\theta^*) \hat{P}_n U'(\theta^*) }{\Phi'}$, where $d_n$ are coefficients that depend on the PDM integrals and $\hat{P}_n$ here are Pauli strings that arise from the transformation of the PDM operator from the second quantized to its qubit operator form. Since we carry out our calculations in an active space, we add to this quantity the frozen core PDM contribution along with the usual nuclear contribution, to obtain the total PDM. For the relativistic calculations, we employ the Dirac--Coulomb Hamiltonian in the Born--Oppenheimer approximation, given by $\hat{H}_{DC}= \sum_k (c \vv{\alpha} \cdot \vv{p_k}+ \beta c^2) +\sum_k V_{nuc}(r_k) + \frac{1}{2} \sum_{k \neq l} \frac{1}{r_{kl}}$, where $\vv{\alpha} =\begin{pmatrix}
0 & \vv{\sigma} \\
\vv{\sigma} & 0
\end{pmatrix}$, 
$\beta =\begin{pmatrix}
\mathbb{I} & 0 \\
0 & -\mathbb{I}
\end{pmatrix}$, 
$\vv{\sigma}$ are the Pauli matrices and $\mathbb{I}$ is the $(2\times2)$ identity matrix. The summations are over the number of electrons. $V_{nuc}(r_k)$ refers to the electron--nucleus potential. We use finite-sized nuclei (Gaussian) for our relativistic calculations. We report all our results in atomic units (au), unless specified otherwise.

We use the following equilibrium bond lengths (in Å): BeH: 1.342 \cite{fazil2018pdm}, MgH: 1.7297 \cite{fazil2018pdm}, CaH: 2.0025 \cite{fazil2018pdm}, SrH: 2.1461 \cite{fazil2018pdm}, BaH: 2.2319 \cite{fazil2018pdm}, and RaH: 2.43 \cite{Fazil2019RaH}. Refs. \cite{fazil2018pdm,Fazil2019RaH} had in turn taken the bond lengths from either experiment or high quality computations available in literature. The one- and two-electron integrals, as well as property integrals, are generated by the DIRAC22 program \cite{DIRAC22}. The VQE-UCCSD statevector computations are performed using Qiskit 0.39.5 \cite{Qiskit2021}, interfaced via modified OpenFermion–Dirac libraries \cite{DIRAC22}. We use uncontracted dyall.v4z basis sets for our simulations \cite{dyall}. First the parameters for VQE computations are initialized with zero initial guesses and parameters in each iteration are updated using the SLSQP (Sequential Least SQuares Programming) optimizer \cite{Kraft1988Seq_QP}. We adopt the Jordan--Wigner mapping scheme throughout and benchmark our VQE-UCCSD PDM values against complete active space configuration interaction (CASCI) calculations, where the Jordan--Wigner-transformed qubit Hamiltonian is diagonalized with the correct particle number using the OpenFermion package~\cite{OpFermion2020McClean}. The Trotter step is set to one throughout. Based on our comparisons with the CASCI values, we also see that this approximation still yields good results for the small problem instances that we have considered (18-qubit active space for simulations, and 6- and 12-qubit calculations on quantum hardware). We also add that changing the Trotter step would not change the number of parameters, but it would drastically increase the circuit depth. 

Relativistic VQE calculations require more quantum resources than their non-relativistic counterparts. The number of circuit evaluations in VQE scales as the product of Hamiltonian terms and iterations (we ignore the number of shots and number of repetitions of an experiment). Relativistic calculations have more non-zero Hamiltonian integrals, resulting in more circuits to evaluate. In the 18-qubit RaH calculation, relativistic VQE requires $12556$ Pauli strings ($47099$ integrals) for the Hamiltonian and $107$ ($162$ integrals) for the PDM operator, compared to $2740$ ($4249$ integrals) and $67$ ($66$ integrals) for the non-relativistic cases, indicating that relativistic VQE is more resource-intensive. 

\section{Results and Discussions}

We now discuss our findings from our 18-qubit VQE simulations (3 occupied and 15 unoccupied spinorbitals). The top panel of Figure \ref{fig:1} illustrates the size of relativistic effects, quantified as \% relativistic effects = $\frac{A_{Rel}-A_{NR}}{A_{Rel}} \times 100 $, where $A$ represents either the ground state energy or the PDM (see Table \ref{SMtable1} of the Supplemental Material for data). It shows that relativistic effects increase significantly from BeH to RaH, with relativity accounting for $7.73\%$ of the ground state energy and $26.47\%$ of the PDM in RaH. Further, VQE-UCCSD and the reference CASCI results disagree by at most $0.47\%$ (for CaH). We note that for BaH (non-relativistic case), the HF calculation initially converged to a metastable spin-doublet state. We obtained the lowest spin-doublet HF state by replacing the initial guess orbitals with HF canonical orbitals from a neighboring geometry. We also note that our PDM results for BeH and MgH are in reasonably good agreement with high-accuracy calculations\cite{fazil2018pdm}, with percentage differences of $2.05\%$ and $3.79\%$ respectively, which may be attributed to our active space sizes and basis set quality. We now examine the role of correlation effects. The bottom panel of Figure \ref{fig:1} shows the \% correlation effects, defined as $\frac{A_{X}-A_{MF}}{A_{X}} \times 100$, for both properties. Here, $A$ can be either the ground state energy or the PDM, and $X$ can be VQE-UCCSD or CASCI. $MF$ is HF for non-relativistic cases and DF for relativistic calculations. The figure shows that the VQE-UCCSD and CASCI results agree remarkably well in predicting correlation energy, within $0.0179\%$. It also shows that VQE-UCCSD and CASCI values are consistent in their prediction of the correlation contribution to the PDM. To understand the interplay between relativistic and correlation effects, we focus on the heaviest RaH molecule. Relativistic effects significantly impact RaH's PDM. Correlation effects decrease its PDM by $1.32\%$ and $1.36\%$ for non-relativistic and relativistic cases, respectively. Conversely, relativity increases the PDM by $26.50\%$ at the mean field level and $26.47\%$ at the correlated level. The combined effect changes the PDM from 1.1306 to 1.5177 au, a total change of $25.51 \%$. \\ 

\emph{Quantum hardware computations: }We compute the active space PDM, henceforth abbreviated as PDM$_{as}$ (the value of the quantity before we manually add the frozen core contributions and nuclear contribution to obtain the total PDM), of SrH in the contracted STO-6G basis in a 12-qubit active space on the IonQ Forte-I device (average 2-qubit gate fidelity: 98.99\%), and also perform 6-qubit active space PDM calculations for SrH and SrF on the IonQ Aria-I device (average 2-qubit gate fidelity: 98.43\%). Since PDM$_{as}$ is what we measure in hardware and thus the errors incurred in hardware reflect changes to this quantity, we only report it in our figures (see Table \ref{SMtable2} of the Supplemental Material for the total PDM values). Despite Forte-I being among the best available commercial quantum computers, achieving high precision on current noisy hardware requires a workflow with several resource reduction techniques, as illustrated in Figure \ref{fig:2}. In particular, sub-figures (b) through (d) present data on the quantum resource reduction, the trade-off in the active space PDM, and the ratio of the number of 2-qubit gates ($2qg$) in the previous step to current step. We outline our workflow below: 

\begin{enumerate}[left=0pt, labelsep=3pt]
\item \textbf{Point group symmetry}: Leveraging the $C_{2v}$ point group symmetry \cite{Cao2022pointgroup} reduces the number of coupled cluster amplitudes, and hence the number of VQE parameters, from 8 to 3 and $2qg$ from 280 to 148 for the 6-qubit case. On the other hand, in the 12-qubit circuit, the number of parameters reduce from 104 to 32 and $2qg$ reduces from 9148 to 2960. 
\item \textbf{Energy Sort VQE (ES--VQE)}: The ES-VQE approach~\cite{Fan2023es} involves carrying out one-parameter VQEs and sorting in descending order the resulting energies, and then carrying out VQE calculations with progressively increasing number of parameters chosen according to the sort. For our purposes, given current-day quantum hardware limitations but also keeping in mind our need to achieve high precision, we pick the parameter with dominant energy contribution for both the 6- and 12-qubit cases (double excitation from (1,3) orbital to (2,5) orbital  and from (1,7) to (5,11) orbital for the 6- and 12-qubit examples respectively). We note that we chose 3 and 5 occupied spinorbitals for the 6- and 12-qubit computations respectively. This process reduces $2qg$ further to 64, with only a $0.33\%$ loss in PDM$_{as}$ for SrH, for the 6-qubit computations. For the 12-qubit case, $2qg$ goes from 2960 to 144, while retaining precision in the PDM$_{as}$ to within $97.65\%$. 
\item \textbf{Pipeline--based optimization}:  A sequence of optimization routines (Qiskit L3 \cite{Qiskit2021}, Pytket \cite{Sivarajah2021tket}, PyZX \cite{kissinger2020Pyzx}, and Qiskit L3 routines applied in that order) further decreases $2qg$ in the 6- and 12-qubit circuits to 37 and 60 respectively. We incur no loss in the PDMs during this step. 
\item \textbf{Reinforcement learning (RL) aided and causal flow preserving ZX--Calculus based approaches}: As a final optimization specifically for the 12-qubit case, we followed the approach described in Ref. \cite{riu2024reinforcementlearningbasedquantum}, where an agent is trained using RL and Graph Neural Networks to apply ZX--Calculus-based graph-theoretic rules for circuit optimization. In this work, we trained our agent with random 5-qubit circuits with 70 \textit{$2qg$}, but with similar gate probabilities as the 12-qubit target circuit. Moreover we defined the reward to focus on \textit{$2qg$} reduction. For the $12q$ case, the agent managed to reduce the \textit{$2qg$} from $60$ to $28$. As a post-processing step, we added the deterministic algorithm from Ref.~\cite{holker2024causal} based on causal flow preserving rules of ZX--Calculus, which in turn reduces the previous \textit{$2qg$} by $2$, to a final $2qg$ of 26. This step is followed by an additional Qiskit L3 pass, which does not affect the number of two-qubit gates, but reduces the number of one-qubit gates further. We incur no loss in PDM after these steps. 

\begin{figure}[ht]
    \setlength{\tabcolsep}{2mm}
        \begin{tabular}{c}
        \hspace{-0.8cm}           
        \vspace{-0.2cm}\includegraphics[height=55mm,width=85mm]{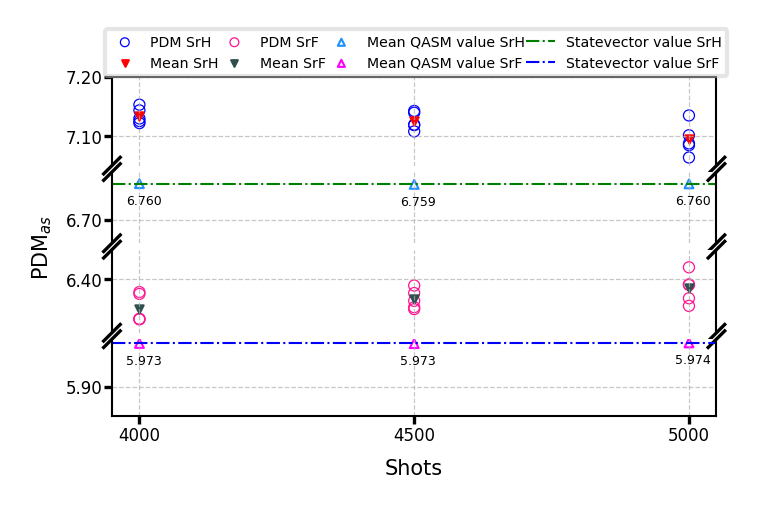} \\ 
        \vspace{-0.1cm} \hspace{0.8cm}
        (a)\\ 
        \hspace{-0.5cm}\vspace{-0.3cm}\includegraphics[height=55mm,width=85mm]{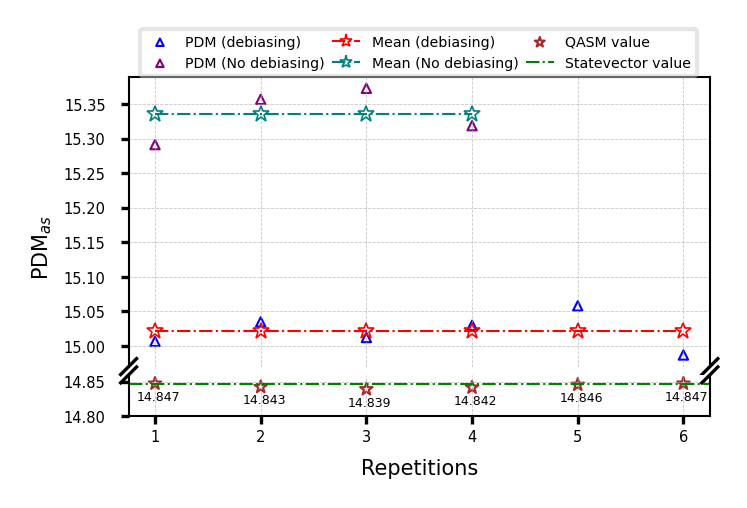}\\
        \vspace{-0.2cm} \hspace{0.8cm}(b)\\
        \end{tabular}
\caption{Active space PDM (PDM$_{as}$) results (in atomic units) for (a) SrH and SrF molecules in a 6-qubit active space with varying shot numbers, averaged over five repeats (each repeat is denoted by a circle, and the average value by a solid variant of `$\grad{}$'), obtained on the 25-qubit IonQ Aria-I hardware after error mitigation and post-selection. The data points denoted with a star symbol give the mean values obtained using the QASM backend on a traditional computer, while the dashed-dotted lines are the statevector results. We note that the main results reported in text take data from the 5000 shots calculation. Sub-figure (b) presents our 12-qubit post-selected results on the IonQ Forte-I device for PDM$_{as}$ of SrH across repetitions (each with 4000 shots), with and without error mitigation. } 
\label{fig:3}
\end{figure} 

\item \textbf{Qubit--wise commuting Pauli groups}: The 6- and 12- qubit PDM computations each involve computing expectation values of 19 and 53 terms respectively, and hence evaluating as many circuits. By partitioning the qubit operator for the PDM into qubit-wise commuting cliques followed by selecting and measuring only the dominant clique (and hence evaluating only one circuit for 6-qubit case and one for the 12-qubit counterpart) for quantum hardware computation, we find that we achieve substantial reduction in the resources with no loss in PDM. 
\item \textbf{Error mitigation and post--selection}: We use the debiasing error mitigation method \cite{maksymov2023symmetry}, followed by post-selecting the particle number conserving bit strings to obtain the PDM$_{as}$ \cite{Bonet2018err_mit, Going2023mol_vqe}. 
\end{enumerate} 

\emph{6-qubit calculation on PDM on IonQ Aria-I: }We discuss results from our quantum hardware computations on the Aria-I device for SrH and SrF molecules. Figure \ref{fig:3}(a) shows the values of PDM$_{as}$, with error-mitigated and post-selected results averaged over 5 repetitions for both the molecules, across different numbers of shots. Considering the computations with 5000 shots as our main result, we see that the percentage fraction difference with respect to statevector results is about $4.94\%$ for SrH and $6.90\%$ for SrF. The 2-qubit gate fidelity was around $98.43\%$. 

\emph{12-qubit calculation on PDM on IonQ Forte-I: }Figure \ref{fig:3}(b) presents our post-selected results for the PDM$_{as}$ of the SrH molecule across different repetitions (each with 4000 shots), with and without error mitigation (debiasing). We find that the mitigated value of the average PDM$_{as}$ that we obtain across 6 repeats is to within $98.81\%$ of the mean QASM value. The figure also indicates that mitigation improves the result by $2.11\%$ relative to the mean QASM value. 

We now report the impact of resource reduction on the convergence of the VQE procedure. Since ES-VQE actually affects the quality of results, we exclude it from the workflow for this analysis. We also exclude RL-based circuit optimization step from the analysis, since the cost of training is prohibitive. We choose the SrH molecule as a representative case, since all of our 18-qubit simulations were carried out with the same size active space (18 spinorbitals). Although our hardware computations were at most 12-qubit calculations, it is worth checking this aspect in a much larger 18-qubit simulation. We compare the circuit with no resource reduction versus the case where a. we include point group symmetry and b. also carry out resource reduction using Qiskit L3 to the input circuit. We fix the optimizer (SLSQP), threshold to convergence (10$^{-6}$), and the initial guess (zero initial guess). We find that the number of function evaluations to convergence changes from 781 to 381 for unoptimized versus optimized circuits, while achieving a reduction in the number of \textit{$2qg$} from 19884 to 8306 and change in energy from $-3178.6336461$ Ha to $-3178.6336453$ Ha. 

\section{Conclusion}

In conclusion, considering the existing gap in the VQE literature concerning the inclusion of relativistic effects and the calculation of properties beyond energies, we implement a relativistic VQE algorithm for calculating molecular permanent electric dipole moments of polar diatomic alkaline earth metal monohydride molecules from BeH to RaH. Benchmarking our 18-qubit VQE simulation results against the complete active space configuration interaction method provides valuable insights into the roles as well as the interplay between correlation and relativistic effects. We achieve a precision of at least $99.72\%$ in capturing relativistic effects for the RaH molecule
with respect to the CASCI value.

Furthermore, we perform high precision quantum hardware computations of the molecular electric dipole moments of the moderately heavy SrH (6- and 12-qubit calculations) and SrF (6-qubit calculation) systems. We employ strategies including point group symmetry, energy sort VQE, pipeline based circuit optimization, reinforcement learning- based ZX--Calculus, and causal flow preserving ZX--Calculus for the 12-qubit computation to compress the circuit by $99.71\%$. We also use debiasing for error mitigation followed by particle number conserving post-selection, to reach a precision of $95.06\%$ and $93.10\%$ (relative to the result obtained by executing the same circuit without noise on a classical device) for the 6-qubit computations on SrH and SrF respectively, and $98.82\%$ for the 12-qubit SrH calculation. This work marks a significant advancement in high-precision relativistic VQE computations, and we expect future progress in quantum hardware to extend these methods to heavier molecular systems and new applications in probing fundamental physics. 

\begin{acknowledgments} 
Calculations were performed on the Rudra cluster (SankhyaSutra Labs) and on the Airawat cluster, CDAC Pune. KS acknowledges support from Quantum Leap Flagship Program (Grant No. JPMXS0120319794) from the MEXT, Japan, Center of Innovations for Sustainable Quantum AI (JPMJPF2221) from JST, Japan, and Grants-in-Aid for Scientific Research C (21K03407) and for Transformative Research Area B (23H03819) from JSPS, Japan. BPD and VSP acknowledge support from MeitY-AWS Braket QCAL project (N-21/17/2020-NeGD, 2022-24). We would like to acknowledge Dr. V. P. Majety (IIT Tirupati) for useful discussions. Shweta acknowledges Dr. Bodhaditya Santra and members of CAQT lab (IIT Delhi) for their support during completion of this project. VSP acknowledges support from Mr. Nishanth Baskaran, Dr. Anjani Priya, Dr. Jean-Christophe and Dr. Daniela for support related to AWS Braket execution. PC acknowledges Mr. Peniel B. Tsemo for useful discussions. 
\end{acknowledgments} 

\bibliography{apssamp}

\newpage

\begin{appendices}

\section*{Supplemental Material}

\renewcommand{\thesection}{s\arabic{section}}
\renewcommand{\theequation}{S\arabic{equation}}
\renewcommand{\thefigure}{S\arabic{figure}}
\renewcommand{\thetable}{S\arabic{table}}

\setcounter{section}{0} 
\setcounter{figure}{0} 
\setcounter{table}{0} 

\begin{table}[t]
    \begin{ruledtabular}
    \caption{\label{SMtable1} Table presenting the ground state energies and PDMs (includes frozen core and nuclear contributions along with the active space contribution) of the considered molecules from different methods. The list of abbreviations used are HF: Hartree--Fock, DF: Dirac--Hartree-Fock, VQE: VQE with UCCSD ansatz,  NR: non-relativistic, Rel: relativistic, CASCI: Complete Active Space Configuration Interaction. Our main results for this work are marked in bold font. The energy is in units of Hartree, whereas the PDM is given in atomic units.
     }
            \begin{tabular}{cccc}
                \textrm{Molecule}&
                \textrm{Method}&
                \textrm{Energy}&
                \textrm{PDM}\\
            \colrule
                BeH & HF & -15.153224 & 0.1137 \\
                 & DF & -15.156052 & 0.1137 \\
                 & \textbf{VQE (NR)} & \textbf{-15.155941} & \textbf{0.0936} \\
                 & \textbf{VQE (Rel)} & \textbf{-15.158768} &  \textbf{0.0937} \\
                 & CASCI (NR) & -15.155949 & 0.0933 \\
                 & CASCI (Rel) & -15.158776 & 0.0933 \\
            
            \colrule
                MgH & HF & -200.157231 & 0.5850 \\
                 & DF & -200.477277 & 0.5861 \\
                 & \textbf{VQE (NR)} & \textbf{-200.159639} & \textbf{0.5288} \\
                 & \textbf{VQE (Rel)} & \textbf{-200.479695} & \textbf{0.5295} \\
                 & CASCI (NR) & -200.159665 & 0.5287 \\
                 & CASCI (Rel) & -200.479722 & 0.5294 \\
                 
            \colrule
                CaH & HF & -677.314268 & 0.8269 \\
                 & DF & -680.265163 & 0.8374 \\
                 & \textbf{VQE (NR)} & \textbf{-677.315076} & \textbf{0.8020} \\
                 & \textbf{VQE (Rel)} & \textbf{-680.265994} & \textbf{0.8111} \\
                 & CASCI (NR) & -677.315085 & 0.8061 \\
                 & CASCI (Rel) & -680.266003 & 0.8150 \\
                 
            \colrule
                SrH & HF & -3132.103015 & 0.9829\\
                 & DF & -3178.633202 & 1.0331 \\
                 & \textbf{VQE (NR)} & \textbf{-3132.103414} & \textbf{0.9650} \\
                 & \textbf{VQE (Rel)} & \textbf{-3178.633645} & \textbf{1.0119} \\
                 & CASCI (NR) & -3132.103417 & 0.9701 \\
                 & CASCI (Rel) & -3178.633649 & 1.0165 \\
                 
            \colrule
                BaH & HF & -7884.116035 & 0.9510 \\
                 & DF & -8136.206375 & 1.0843\\
                 & \textbf{VQE (NR)} & \textbf{-7884.116240} & \textbf{0.9400} \\
                 & \textbf{VQE (Rel)} & \textbf{-8136.206603} & \textbf{1.0715} \\
                 & CASCI (NR) & -7884.116240 & 0.9446 \\
                 & CASCI (Rel) & -8136.206609 & 1.0755 \\
                 
            \colrule
                RaH & HF & -23094.880586 & 1.1306\\
                 & DF & -25028.736164 & 1.5383\\
                 & \textbf{VQE (NR)} & \textbf{-23094.880796} & \textbf{1.1159} \\
                 & \textbf{VQE (Rel)} & \textbf{-25028.736452} & \textbf{1.5177} \\
                 & CASCI (NR) & -23094.880796 & 1.1203 \\
                 & CASCI (Rel) & -25028.736502 & 1.5180 \\
                 
    \end{tabular}
    \end{ruledtabular}
\end{table}

\begin{table}[t]
    \begin{ruledtabular}
    \caption{\label{SMtable2} 
    Table presenting the PDMs (includes frozen core and nuclear contributions along with the active space contribution) in the relativistic regime of the molecules considered for hardware execution. The `$1 \theta^*$' refers to the fact that the UCCSD computation is carried out with only one parameter (the most dominant one). The PDM is given in atomic units.
     }
            \begin{tabular}{ccc}
                \textrm{Molecule}&
                \textrm{Method}&
                \textrm{PDM}\\
            \colrule
                SrH ($12q$) & DF & 1.3695 \\
                 & UCCSD & 0.9828 \\
                 & CASCI & 0.9729  \\
                 & \textbf{UCCSD}($1 \theta^*$)  &  1.3402\\
                 & \textbf{Hardware} & 1.1655 \\
            
            \colrule
                SrH ($6q$) & DF & 1.3695  \\
                 & UCCSD & 1.3410 \\
                 & CASCI & 1.3247  \\
                 & \textbf{UCCSD}($1 \theta^*$)  & 1.3636 \\
                 & \textbf{Hardware} & 1.0293 \\
                 
            \colrule
                SrF ($6q$) & DF & 1.6145 \\
                 & UCCSD & 1.5910  \\
                 & CASCI & 1.5755  \\
                 & \textbf{UCCSD}($1 \theta^*$)  & 1.6099\\
                 & \textbf{Hardware} & 1.1978 \\
    \end{tabular}
    \end{ruledtabular}
\end{table}

\begin{table*}[hbt!]
    \begin{ruledtabular}
    \caption{\label{SMtable3} Table presenting the clique-wise contribution to PDM at HF and correlated levels for SrH (6- and 12-spinorbital active space denoted as `6q' and `12q' respectively) and SrF (6-spinorbital space) with 1-parameter VQE statevector calculation. For this case, the entire contribution to the PDM, both at mean field level and correlated level, comes from Pauli strings containing only $I$ and $Z$. For a given molecule, each row is a clique (qubit-wise commuting), and lists the number of Pauli words in each clique.  }
        \begin{tabular}{ccccc}
                \textrm{Molecule}&
                \textrm{Terms}&
                \textrm{${\bra{\Phi_0}\hat{D}\ket{\Phi_0}}$}&
                \textrm{${_r\bra{\Psi}\hat{D}\ket{\Psi}}_r$}& 
                \textrm{Correlation in PDM (a.u.)} \\
            \colrule
                SrH (6q) & {\textbf{IIIIII}, IIIIIZ, IIIIZI, IIIZII, IIZIII, IZIIII, ZIIIII} & 6.75406 & 6.75997 & 0.00591  \\
            
                 & {IIIIYY, IIIYYI, IYYIII, YYIIII} & 0 & 0 & 0\\
            
                 & {IIIIXX, IIIXXI, IXXIII, XXIIII} & 0 & 0 & 0 \\
                 & {IIIYZY, YZYIII} & 0 & 0 & 0 \\
                 & {IIIXZX, XZXIII} & 0 & 0 & 0 \\
            \colrule
                SrF (6q) & {\textbf{IIIIII}, IIIIIZ, IIIIZI, IIIZII, IIZIII, IZIIII, ZIIIII} & 5.96841 & 5.97296 & 0.00455  \\
                 & {IIIIYY, IIIYYI, IYYIII, YYIIII} & 0 & 0 & 0\\
                 & {IIIIXX, IIIXXI, IXXIII, XXIIII} & 0 & 0 & 0 \\
                 & {IIIYZY, YZYIII} & 0 & 0 & 0 \\
                 & {IIIXZX, XZXIII} & 0 & 0 & 0 \\
            \colrule
                SrH (12q) & {\textbf{IIIIIIIIIIII}, IIIIIIIIIIIZ, IIIIIIIIIIZI} &  &  &  \\ 
                &{IIIIIIIIIZII, IIIIIIIIZIII, IIIIIIIZIIII} & & & \\ 
                &{IIIIIIZIIIII, IIIIIZIIIIII, IIIIZIIIIIII} & & & \\ 
                &{IIIZIIIIIIII, IIZIIIIIIIII, IZIIIIIIIIII} & & & \\ 
                &{ZIIIIIIIIIII} & 14.81738 & 14.84663 & 0.0292 \\
                \\
            
                & {IIIIIIIIIIYY, IIIIIIIIIYYI, IIIIIIIIYYII} &  &  &  \\  
                &{IIIIIIYZYIII, IIIIYYIIIIII, IIIYYIIIIIII} & & & \\ 
                &{IIYYIIIIIIII, YZYIIIIIIIII} & 0  & 0  & 0 \\
                \\

                &{IIIIIIIIIIXX, IIIIIIIIIXXI, IIIIIIIIXXII} &  & & \\  
                
                &{IIIIIIXZXIII, IIIIXXIIIIII, IIIXXIIIIIII} & & & \\ &{IIXXIIIIIIII, XZXIIIIIIIII} & 0 & 0 & 0 \\
                \\

                & {IIIIIIIIIYZY, IIIIIIYZZYII, IIIYZYIIIIII, YZZYIIIIIIII} & 0 & 0 & 0 \\
                \\

                & {IIIIIIIIIXZX, IIIIIIXZZXII, IIIXZXIIIIII, XZZXIIIIIIII} & 0 & 0 & 0 \\
                \\

                & { IIIIIIIIYZZY, IIYZZYIIIIII} & 0 & 0 & 0 \\
                \\

                & {IIIIIIIIXZZX, IIXZZXIIIIII} & 0 & 0 & 0 \\
                \\

                & {IIIIIIYZZZZY, YZZZZYIIIIII} & 0 & 0 & 0 \\
                \\

                & {IIIIIIXZZZZX, XZZZZXIIIIII} & 0 & 0 & 0 \\
                \\

                & {IIIIIIIIYZYI, IIYZYIIIIIII} & 0 & 0 & 0 \\
                \\

                & {IIIIIIIIXZXI, IIXZXIIIIIII} & 0 & 0 & 0 \\
                \\

                & {IIIIIIYZZZYI, YZZZYIIIIIII} & 0 & 0 & 0 \\
                \\

                & {IIIIIIXZZZXI, XZZZXIIIIIII} & 0 & 0 & 0 \\
                \\

        \end{tabular}
\end{ruledtabular}
\end{table*}

\end{appendices}

\end{document}